# Algorithms for determining resistances in quantum Hall annuli with *p-n* junctions

Chieh-I Liu[1,2], Dominick S. Scaletta[3], Dinesh K. Patel[1,4], Mattias Kruskopf[1,5,6], Antonio Levy[1], Heather M. Hill[1], and Albert F. Rigosi[1]

[1]Physical Measurement Laboratory, National Institute of Standards and Technology (NIST), Gaithersburg, MD 20899, United States
[2]Department of Chemistry and Biochemistry, University of Maryland, College Park, MD 20742, United States
[3]Department of Physics, Mount San Jacinto College, Menifee, CA 92584, United States
[4]Department of Physics, National Taiwan University, Taipei 10617, Taiwan
[5]Joint Quantum Institute, University of Maryland, College Park, MD 20742, United States
[6]Electricity Division, Physikalisch-Technische Bundesanstalt, Braunschweig 38116, Germany

E-mail: afr1(at)nist.gov

**Abstract**

Just a few of the promising applications of graphene Corbino *pn*J devices include two-dimensional Dirac fermion microscopes, custom programmable quantized resistors, and mesoscopic valley filters. In some cases, device scalability is crucial, as seen in fields like resistance metrology, where graphene devices are required to accommodate currents of the order 100 μA to be compatible with existing infrastructure. However, fabrication of these devices still poses many difficulties. In this work, unusual quantized resistances are observed in epitaxial graphene Corbino *p-n* junction devices held at the $\nu = 2$ plateau ($R_\mathrm{H} \approx 12906\ \Omega$) and agree with numerical simulations performed with the LTspice circuit simulator. The formulae describing experimental and simulated data are empirically derived for generalized placement of up to three current terminals and accurately reflects observed partial edge channel cancellation. These results support the use of ultraviolet lithography as a way to scale up graphene-based devices with suitably narrow junctions that could be applied in a variety of subfields.

## 1. Introduction

Graphene and all devices fabricated from it have been studied extensively since its discovery [1-4]. Under strong magnetic flux densities leading to filled Landau levels, graphene exhibits fixed resistances that take the form $\frac{1}{2(2n+1)}R_\mathrm{K}$, where $R_\mathrm{K} = \frac{h}{e^2}$ and is labelled as the von Klitzing constant, *n* is an integer, *h* is the Planck constant, and *e* is the elementary charge. Conventional *p-n* junction (*pn*J) Hall devices may also exhibit a variety of ratios of the von Klitzing constant while in the quantum Hall regime [5-18]. Furthermore, similar phenomena have been observed in devices with a Corbino geometry [19-25]. When coupled with the commercial necessity of scaling graphene devices, applications involving millimeter-scale fabrication have the potential to provide solutions in a number of fields, notably those that focus on problems in quantum phenomena in other 2D materials [26-30], quantum Hall metrology [31-41], and electron optics [42-45].

The first question that may come to mind regards how such devices could be applied specifically to various problems. Applications of these Corbino *pn*J devices include the possible construction of more sophisticated two-dimensional Dirac fermion microscopes that rely on large-scale junction interfaces [46], custom programmable quantized resistors [47], and mesoscopic valley filters [21]. The scalability is crucial for some of these applications. For instance, in resistance metrology, graphene devices are required to accommodate currents of the order 10 μA and above (modern-day usage may even exceed 100 μA) in order to ensure compatibility with existing infrastructure [31, 37, 40].

Two difficult steps in successfully fabricating millimeter-scale *pn*J devices include the following: (1) uniformly doping large-area regions on epitaxial graphene (EG) such that it may exhibit both *p*-type and *n*-type behavior and (2) ensuring adequate junction narrowness to enable Landauer-Büttiker edge channel propagation and equilibration [5-9, 48-53]. For the first case, common nanodevice fabrication practices such as using a top-gate are unable to be used due to an increasing probability of current leakage through the gate with lateral size. Furthermore, such typical practices are time-consuming when scaled up beyond the micron level. Comparisons on other fabrication techniques are provided in the Supplementary Material.

Other further specific applications of interest to those exploring quantum Hall transport may include the utilization of *pn*J devices for accessing different quantized resistances or the repurposing of Corbino geometries for quantum Hall devices. In the latter case, not much has been reported regarding how a periodic boundary condition affects measured quantized resistances.

Recent studies show that the parameter space for quantized resistances opens up signficantly when using several terminals as sources or drains [54-57]. In only one of those cases, Corbino *pn*J devices were used, but mostly as a proof of principle for a more complex quantum dartboard device [57]. The empirical understanding of how these values are obtained is still lacking.

This work reports details on the millimeter-scale fabrication of EG Corbino *pn*J devices and subsequent measurements of those devices in the quantum Hall regime to understand how periodic boundary conditions on edge channel currents affect quantized resistances. The data were compared with LTspice current simulations [58-59], and both were then used as the basis for deriving empirical formulae for the generalized case of using two or three current terminals of either polarity with any arbitrary configuration.

Overall, these experiments further validate two endeavors: (1) fabrication of scalable of *pn*J devices and their versatility in circuits (2) flexibility in device fabrication by transforming devices with Corbino geometries into ones that permit the flow of edge channel currents between the outer and inner edges [21, 52].

## 2. Experimental and Numerical Methods

*2.1 Graphene growth and device fabrication*

EG was grown on a 2.7 cm by 2.7 cm SiC square that was diced from a 4*H*-SiC(0001) wafer (CREE) [see Notes]. The procedures for cleaning and treating the wafer before the growth are detailed in other works [32, 35, 54]. One crucial element to obtaining high-quality growth with limited SiC step formation was the AZ5214E solution, a polymer which has been shown to assist in homogenous sublimation [60]. The growth was performed at 1900 °C in an argon environment using a resistive-element furnace from Materials Research Furnaces Inc. [see Notes] with graphite-lining and heating and cooling rates of about 1.5 °C/s.

Samples were inspected after growth with confocal laser scanning and optical microscopy to verify monolayer homogeneity [61]. For fabrication processes, it was important to protect the EG from photoresists and organic contamination, and this was achieved by depositing Pd and Au layers [32, 35]. For improved cryogenic contact resistances, EG was contacted with pads composed of NbTiN, a superconducting alloy with a $T_c$ of about 12 K at 9 T [34, 41]. All EG Corbino *pn*J devices underwent functionalization treatment with $Cr(CO)_6$, which sublimates in a furnace and decomposes into $Cr(CO)_3$ and bonds itself to the EG surface [62-65]. This treatment both provides uniformity along the millimeter-scale devices and reduces the electron density to a low value of the order $10^{10}$ cm$^{-2}$, thus enabling a greater control of the latter by annealing [66].

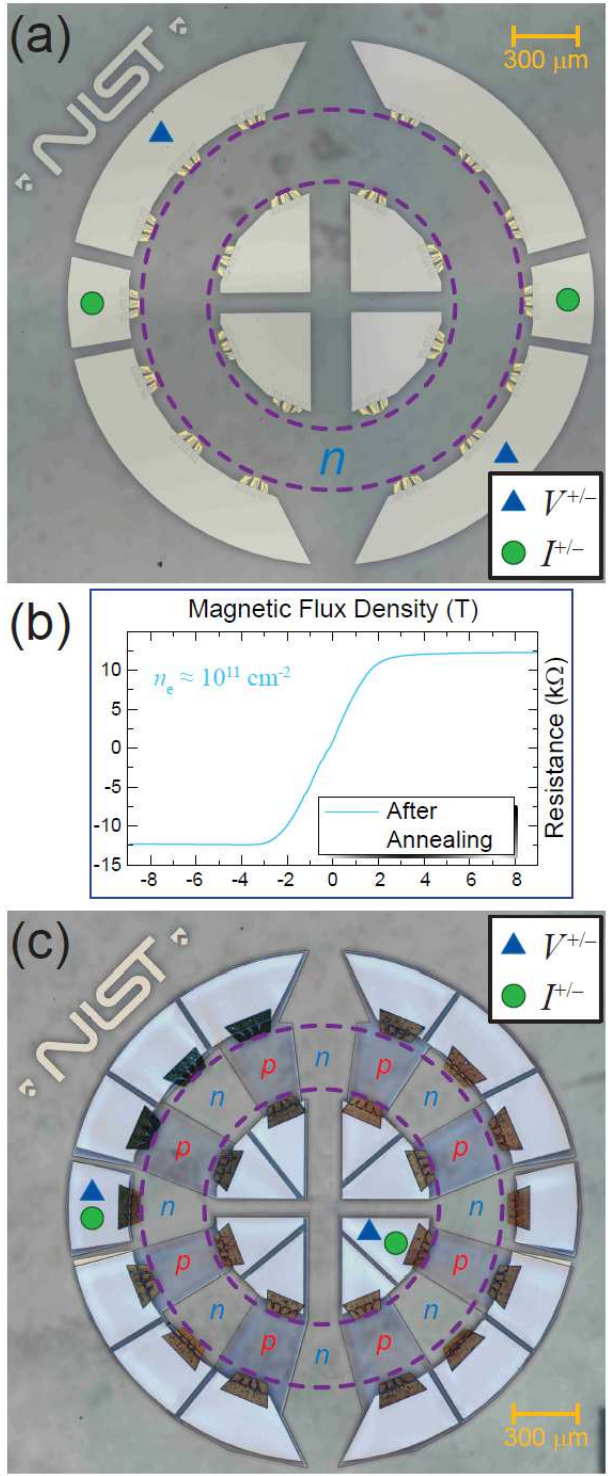

Fig. 1. (a) Optical image of an example Corbino device assigned as a control to determine the necessary annealing conditions for obtaining suitable *n*-type regions. Purple dashed rings indicate the bounds of the epitaxial graphene. Green dots and blue triangles indicate current and voltage terminals, respectively, for the corresponding Hall measurement shown in (b) Optical image of final experimental device containing 16 distinct and alternating *n*-type and *p*-type regions. Green dots and blue triangles are shown for an example configuration (in this case, a two-terminal measurement).

For both the control and experimental devices, intended *n*-type regions were protected by S1813 photoresist. Keeping control devices aside, ultraviolet photolithography was then used to remove S1813 from regions intended for *p*-type adjustment. PMMA/MMA was deposited as a mediation layer for ZEP520A, a polymer with photoactive properties. The latter enables graphene to become *p*-type (near $4 \times 10^{11}$ cm$^{-2}$) upon exposure to an external ultraviolet lamp (254 nm) – see Supplementary Material [54, 67]. Regions still protected by S1813 did not undergo significant electron density shifting but still required an annealing process of approximately 25 min (at 350 K) to shift the electron density to about $10^{11}$ cm$^{-2}$.

To verify that the devices are properly adjusted to the desired electron density, two types of measurements were required. For the control device in Fig. 1 (a), a simple Hall measurement was performed after annealing using the green dots as the current terminals and the blue triangles as the voltage terminals. An example result is shown in Fig. 1 (b), where the electron density has been successfully shifted from low values neighboring the Dirac point to around $10^{11}$ cm$^{-2}$. This electron density is sufficient to see the quantized plateau at ν = 2, which, for the case of using epitaxial graphene, exhibits a stable plateau for a large range of magnetic flux densities. This stability, labelled as a pinning of the ν = 2 Landau level state and characterized by edge channels of opposite chirality, has been attributed to field-dependent charge transfer between the SiC surface and the graphene layer [33].

The second measurement is explained in more detail in the Supplementary Material. In essence, a traditional Hall bar with a *pn*J was fabricated using identical steps. Simple Hall data in the intended *p*-type region was collected to show the electron (or hole, in this case) density after the exposure to the ultraviolet lamp. The annealing does shift *p*-type regions slightly closer to the Dirac point, but the density remains well within the order $10^{11}$ cm$^{-2}$. Additional data from monitoring the carrier density during the photochemical gating process are also shown in the Supplementary Material.

Though these two measurements are direct ways of obtaining the electron density, an indirect way of validating device functionality is to assess the agreement between two- and three-terminal simulations and corresponding experimental data. These analyses are part of the core of this work and will be presented in the next section.

### 2.2 Definitions for empirical framework

Before continuing, one major assumption of the more specific framework below is that all regions are quantized at the ν = 2 plateau. That said, this framework may be reformulated to accurately reflect the conditions of any quantum Hall *pn*J system, including conditions whereby some regions exhibit other plateaus such as the ν = 6 plateau. Now, to thoroughly investigate the large parameter space of quantized resistances subject to periodic boundary conditions, multiple current terminals must be used. One of the goals of this work is to develop an empirical framework for calculating the effective quantized resistance of the circuit shown in Fig. 2. Definitions for that framework include: (1) *N*, the total number of terminals, (2) $q_{N-1}$ and (3) $q_{N-1}^L$ are the *coefficients of effective resistance* (CER) for the cases with (Corbino device) and without (traditional Hall bar device) periodic boundary conditions, respectively, (4) $n_j$, where *j* can be either 1 or 2 and is used to label the number of junctions between two terminals, (5) *M*, the number of distinct regions in the Corbino *pn*J device (must be an even, positive integer), and (6) $n_x$, where $n_x = M - n_1$ for two-terminal circuits and $n_x = M - n_1 - n_2$ for three-terminal circuits.

For greater clarity, refer to the schematics in Fig. 2 (a) and (b), which represent the device in Fig. 1 (c) and are topologically identical (the actual schematic for LTspice simulations is accurately reflected by (b)). The experimental device has *M* = 16. The *pn*J circuit contains a total of 3 terminals (*N* = 3), with the voltage always being measured between points *A* and *B* (green squares). This measurement yields a quantized resistance of the form $R_{AB} = q_{N-1}R_H$, where $R_H$ is the Hall resistance at the ν = 2 plateau ($R_H \approx 12906\ \Omega$). The CER ($q_{N-1}$) can be represented as a either an integer or a fraction.

This work focused on varying the locations of the two (*N* = 2) or three (*N* = 3) current terminals, arbitrary in both position along the Corbino device and placement within the outer or inner circumference. The next step was to determine the best way of identifying $n_1$ (and $n_2$ for the *N* = 3 case). These determinations and corresponding simulations will be shown and discussed in the results section.

### 2.3 LTspice simulations

The electronic circuit simulator LTspice was used for predicting the electrical behavior of the graphene Corbino *pn*J devices. The circuit comprised interconnected *p*-type and *n*-type quantized regions that were modeled either as ideal clockwise (CW) or counterclockwise (CCW) *k*-terminal quantum Hall effect elements. The terminal voltages and currents, represented as $e_m$ and $j_m$, are related by $R_H j_m = e_m - e_{m-1}$ ($m = 1, ..., k$) for CW elements and $R_H j_m = e_m - e_{m+1}$ for CCW elements. The circuit's behavior at *A* and *B* (Fig. 2 (b)) could only be modeled for one polarity of magnetic flux density per simulation. For a positive *B*-field, an *n*-doped (*p*-doped) graphene device was modeled by a CW (CCW) element, whereas, when *B* is negative, a CWW (CW) element was used.

## 3. Results

### 3.1 Interpreting simulation trends (N = 2)

Simulations were first carried out for the *N* = 2 case (which, by default, is one positive and one negative current terminal). By keeping the positive terminal (source) fixed on an arbitrary terminal on the outer circumference of the device, and by moving the negative terminal (drain) along both the outer and inner circumference, the resulting CERs (labelled $q_1$) were simulated as a function of junction number $n_1$ between the two terminals, for several devices containing different numbers of total regions *M*. These results are summarized in Fig. 3 (a).

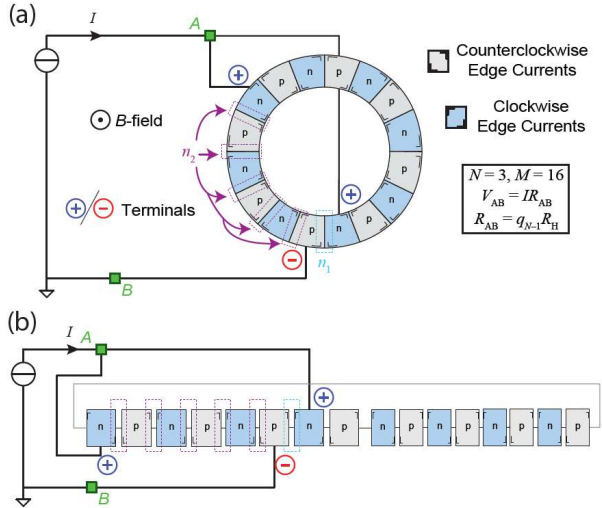

Fig. 2. (a) Schematic of the graphene Corbino *pn*J device from Fig. 1 (c) is shown as part of a circuit intended to exhibit many quantized resistances. In this case, two positive current terminals were used (with each the outer and inner ring hosting one terminal) and one negative terminal was used (outer ring). (b) A topologically identical schematic of the device is shown and accurately reflects the configuration of the quantum Hall elements (*n*-type and *p*-type regions) in the LTspice simulation.

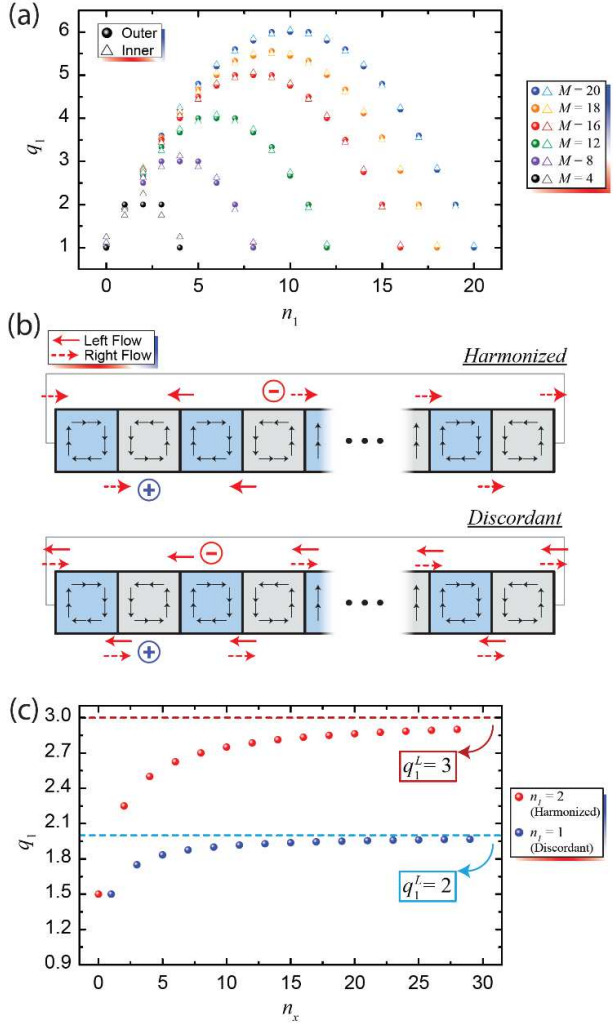

Fig. 3. (a) Data representing the simulated CERs of the two-terminal measurements for Corbino *pn*J devices of varying number of distinct regions, $M$. (b) The illustrations shown here exemplify a physical interpretation for why an alternating behavior is observed in the simulations whereby the negative terminal is moved along the outer or inner circumference. (c) The two configurations in (b) are simulated for varying $n_x$, with the results providing insight into how one may express a general formula to calculate the CER of an arbitrary $N$ = 2 case.

In the case where a positive terminal is held on the outer circumference of the device and a negative terminal is moved along the outer circumference, a parabolic trend appears to form having an intuitive symmetry like the device itself. However, alternating behavior was observed along this parabolic trace. Similarly, when the negative terminal is instead simulated along the inner circumference, a parabolic trend is also seen with alternating behavior. The combination of both, seen in Fig. 3 (a), suggests that two parabolic trends actually exist, with one of them taking on slightly lower values than the other.

There are two consistent physical pictures that arise from the periodic boundary conditions, and these may provide insight into how to interpret the observed alternating behavior. Consider the cases shown in Fig. 3 (b). With the condition that current flows only if it eventually terminates on a positive terminal, then in one case, current is allowed to flow along the edges unimpeded by any other flow. Let us label this as a *harmonized* configuration. The second case involves current flow that impedes itself in several regions of the device. There are special cases (within the $N$ = 3 configuration) where this impeding leads to outright cancellation, enabling the device to emulate a traditional Hall bar with several *pn*Js. All instances of currents appearing to self-impede in this picture may be labelled as *discordant*.

Separating configurations as harmonized or discordant allows the data in Fig. 3 (a) to be fit to a parabola exactly. In doing so, one may parameterize the problem for arbitrary devices and terminal placements. For this analysis, since $n_1$ is symmetric, one may choose $n_1$ to be the smaller spacing between the two terminals, leaving the larger one to be $n_x = M - n_1$. In the limit where $n_x \to \infty$, the periodic boundary condition is effectively lifted, giving us a CER of $q_1^L$, which may be calculated for the traditional Hall bar case [56]. By simulating the CERs ($q_1$) as a function of $n_x$ (see Fig. 3 (c)), a logistic function known as the Hill-Langmuir equation may be used to fit the curves exactly:

$$q_1(n_x) = B + \frac{A - B}{1 + \left(\frac{n_x}{x_0}\right)^p} = \frac{Bn_x + Ax_0}{n_x + x_0}$$

(1)

The parameters in Eq. (1) can be interpreted as meaningful quantities (with $p = 1$). With the limiting case described earlier, $B = q_1^L$, and as $n_x \to 0$, $q_1 = A \equiv q_1^{(0)}$. For all $N$ = 2 configurations, $x_0 = n_1$. Furthermore, with the relation $q_1^L = n_1 + 1$ [56], a function of $n_1$ can be expressed:

$$q_1(n_x \to n_1) = \frac{(n_1 + 1)(M - n_1) + q_1^{(0)} n_1}{M}$$

(2)

In Eq. (2), $q_1^{(0)}$ can be interpreted as the initial condition for a fixed $n_1$ (and $n_x = 0$). It takes on a single value for all harmonized and discordant (within $N$ = 2) – either $\frac{(n_1+1)}{n_1}$ or $\frac{n_1}{n_1} = 1$, respectively. This distinction contributes to the observed separation of the two similar parabolas seen in Fig. 3 (a) and expressed exactly in Eq. (2).

### 3.2 Comparing experimental data to corresponding simulations ($N$ = 2)

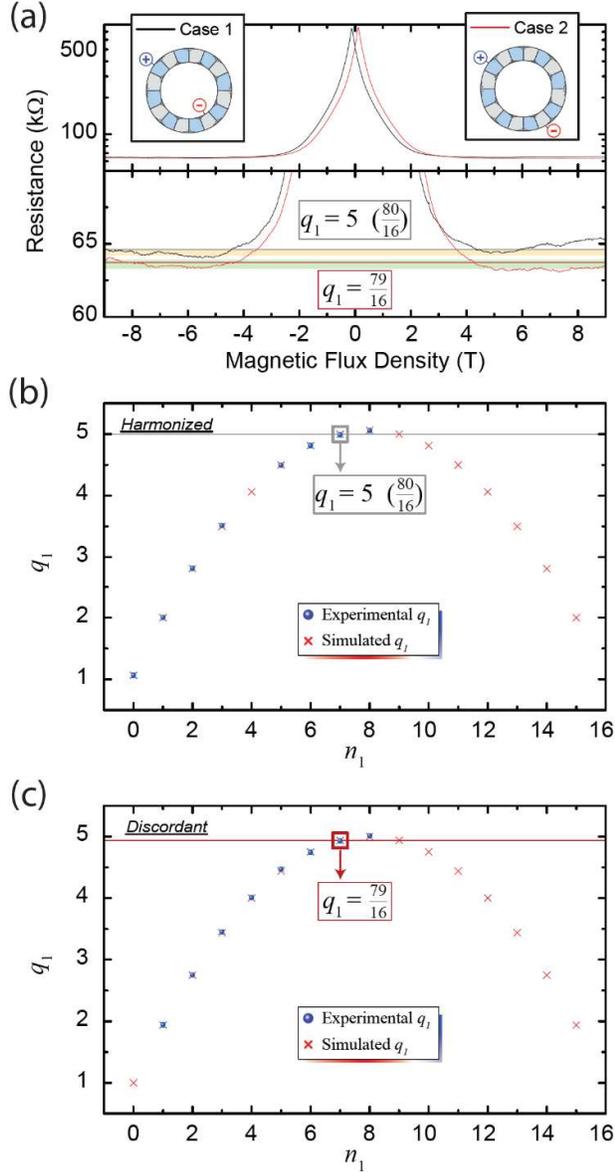

Fig. 4. (a) Magnetoresistance measurements were performed for a variety of $N = 2$ configurations on the device shown in Fig. 1 (c). Two example magnetic flux density sweeps are shown in black and red for the harmonized and discordant case of $n_1 = 7$, respectively. The thin gray and dark red lines are the simulated quantized values, and the shaded gold and green regions are the 1σ uncertainty regions of the respective experimental values. (b) The CERs were simulated (red X) and compared with experimental data (blue points) in harmonized cases as a function of $n_1$. (c) CERs were simulated and compared with experimental data in discordant cases as a function of $n_1$. Error bars (same 1σ uncertainty as exemplified in (a)) are shown in light blue and fall within the size of the blue points in most cases.

To assess the validity of Eq. (2), measurements were performed at a temperature of 1.6 K, with a current of 1 μA, on the device shown in Fig. 1 (c) ($M = 16$). The Supplementary Material also includes information about the mobility of the devices, which range from 3000 cm$^2$V$^{-1}$s$^{-1}$ and 5000 cm$^2$V$^{-1}$s$^{-1}$ for both region types. Recall that regarding edge channel dynamics in a bipolar graphene *pn*J, the quantized states exhibited by the ν = 2 plateau circulate in opposite directions and merge to form a parallel edge channel at the junction. These channels, as mentioned in Ref. [50], supply particles at the junction from both reservoirs. After particles jointly propagate along the interface and to the device boundary, they return to their respective regions. Resistance quantization was explained by mode-mixing at the junction, with the idea that regardless of reservoir, all incoming charges had the same probability of crossing the junction [50]. For information regarding quantum shot noise and Fano factor calculations, please see the Supplementary Material. Overall, these dynamics manifest themselves as a quantized resistance across the junction and can be treated as a circuit element in LTspice.

In Fig. 4 (a), two example measurements taken between ± 9 T are shown in black and red for the harmonized and discordant case of $n_1 = 7$, respectively. For Case 1 (black line), a thin gray line is used to mark the simulated CER of 5, and a shaded gold region marks the 1σ uncertainty of the experimental average, as calculated by the whole range excluding -5 T to 5 T. For Case 2 (red line), a dark red line is used to mark the simulated CER of $\frac{79}{16}$, with a corresponding experimental uncertainty range shaded in green. The simulated values fall within the error of the experimentally-obtained values.

The CERs were calculated with Eq. (2) for the $M = 16$ device and are shown in Fig. 4 (b) and (c). The calculations agreed exactly with the simulations, as expected. Both the calculations and simulations are represented by a red 'X' and were compared with experimental data, represented by blue points, for both harmonized and discordant cases. The error bars are shown in light blue, with many falling within the size of the experimental data points. The same gray and red lines from Fig. 4 (a) are shown, along with a box surrounding the relevant data points. These markers enhance the clarity of the difference between the harmonized and discordant cases. The agreement between the experiment and calculated CERs supports the validity of Eq. (2) for all $N = 2$ configurations.

### 3.3 Interpreting simulation trends (N = 3)

Simulations were next carried out for the $N = 3$ case (two terminals of a single polarity and one terminal of opposite polarity). The CERs (now labelled $q_2$) of numerous arbitrary configurations were again simulated as a function of junction number $n_x = M - n_1 - n_2$, where $n_x$ is defined between the two like-polar terminals. The other two numbers $n_1$ and $n_2$ describe the junction number between the two opposite-polarity pairs, with $n_1$ being the smaller number to be consistent with the traditional Hall bar case [56].

Two example simulation sets are shown in Fig. 5 (a), with both sets having $n_1 = 1$ and $n_2 = 3$. The number of regions $M$ was modulated, allowing one to model $q_2(n_x)$. Both the harmonized and discordant cases were modeled exactly to the Hill-Langmuir equation, and the limiting case of $n_x \to \infty$ revealed again that $q_2 \to q_2^L$, which can be calculated [56]. In the case of Fig. 5 (a), $q_2^L = \frac{8}{5}$, and this value is marked by a dashed line. Additionally, $q_2^{(0)}$ is marked for both cases. The two values at $n_x = 12$ are simulated values with corresponding experimental data shown in the first cases of Fig. 5 (b) and (c).

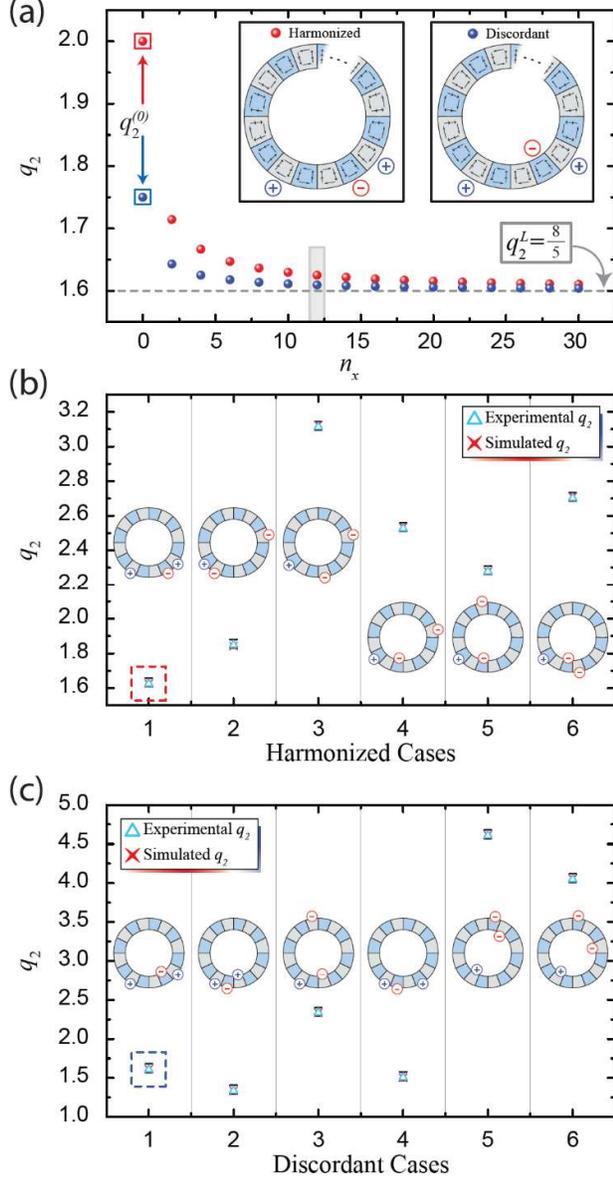

Fig. 5. (a) Simulations for the two shown configurations were performed while varying $n_x$. (b) Experimental data for a variety of harmonized and (c) discordant cases are compared with their simulated counterparts (and verified again with Eq. (4)). The exact configuration is depicted for each case, and error bars indicate 1σ uncertainty and are of similar size to the light blue triangles (experimental data points) in most cases.

By rewriting Eq. (1) and (2), one may more clearly see the iterative nature of the formula that will describe all $N = 3$ cases. Recall that for all $N = 2$ cases:

$$q_1(n_1) = \frac{q_1^L(M - n_1) + q_1^{(0)} n_1}{(M - n_1) + n_1}$$
(3)

Here, the only term that changes for harmonized or discordant cases is $q_1^{(0)}$. For all cases in $N = 3$, the parameter $x_0 = \frac{n_1 + n_2}{n_1 + n_2 + 1}$, and the general CER formula becomes:

$$q_2(n_1, n_2) = \frac{q_2^L(M - n_1 - n_2) + q_2^{(0)} x_0}{(M - n_1 - n_2) + x_0}$$
(4)

And again, the difference between harmonized and discordant cases is embedded in the term $q_2^{(0)}$, which takes on the values $\frac{(n_1+1)(n_2+1)}{n_1+n_2}$ or $\frac{n_1+n_2+n_1 n_2}{n_1+n_2}$, respectively (see Supplementary Material for more details on how these values were determined).

### 3.4 Comparing experimental data to corresponding simulations (N = 3)

To verify Eq. (4), data were collected from several $N = 3$ cases. Six example harmonized and discordant cases are shown in Fig. 5 (b) and (c), respectively. Each experimental data point (light blue triangle) very nearly overlays with its corresponding simulation (red 'X'), and the simulations match the calculations exactly. Additionally, each point is accompanied by an illustration of each configuration. The error bars, in a darker shade of blue, indicate 1σ uncertainty and have a similar size as the experimental data points in most cases. The exact CERs for all presented experimental data are listed in the Supplementary Material. The agreement within uncertainty with simulations demonstrates promise that these large-scale devices can be fabricated with excellent functionality.

### 4. Conclusion

This work reports the successful fabrication of millimeter-scale graphene Corbino *pn*J devices and corresponding measurements of such devices in the quantum Hall regime to understand how the edge channel currents resulting from being in the ν = 2 plateau, manifesting as quantized effective circuit resistances, are affected by periodic boundary conditions. Experimental data were compared with results from LTspice current simulations. Furthermore, empirical formulae were derived for the case of using two or three current terminals of arbitrary configuration. Overall, these experiments have validated that these scalable *pn*J devices are versatile in how they are implemented in circuits and that using Corbino geometries to permit edge channel current flow between the outer and inner edges offers another adjustable parameter for quantum electrical circuits.

**Acknowledgements and Notes**

The work of DKP at NIST was made possible by C-T Liang of National Taiwan University. The authors would like to express thanks to S Payagala and X Wang for their assistance in the NIST internal review process.

Commercial equipment, instruments, and materials are identified in this paper in order to specify the experimental procedure adequately. Such identification is not intended to imply recommendation or endorsement by the National Institute of Standards and Technology or the United States government, nor is it intended to imply that the materials or equipment identified are necessarily the best available for the purpose.